\documentclass[preprint]{revtex4}
\usepackage{amsmath, amsfonts, amssymb,graphicx, dcolumn}

\begin{document}

\title{Robust Multi-Partite Multi-Level Quantum Protocols}
\author{Hideomi Nihira}
\email{nihira@optics.rochester.edu}
\author {C. R. Stroud, Jr.}
\affiliation{The Institute of Optics, University of Rochester,
Rochester, NY 14627, USA}
\date{\today}

\begin{abstract}
We present a tripartite three-level state that allows a secret
sharing protocol among the three parties, or a quantum key
distribution protocol between any two parties. The state used in
this scheme contains entanglement even after one system is traced
out. We show how to utilize this residual entanglement for quantum
key distribution purposes, and propose a realization of the scheme
using entanglement of orbital angular momentum states of photons.
\end{abstract}

\maketitle

\section{Introduction}

Here we present a simple three-level tripartite quantum protocol
that can be generalized to a N-level N-partite scheme.  The
initial state that the three parties share can be used for both
quantum secret sharing protocol or a BB84-like protocol between
any two parties. Although BB84 is a protocol in which Alice and
Bob perform a measurement on the same particle, making separate
measurements on two entangled particles shared between Alice and
Bob can also produce perfectly correlated measurement outcomes. In
this manner a BB84-like scheme can be employed for entangled
states.  The interesting aspect of the proposed state is that,
unlike the $|GHZ\rangle _{N}$ state, the reduced density matrix of
any two particles still contains some entanglement, and perfectly
correlated measurements can be made in the reduced space thus
making the protocol robust against particle loss.  We also show
how this scheme can be realized using entangled orbital angular
momentum states of light.

    The original cryptographic protocol introduced by Bennett and
Brassard \cite{Bass1} generated a secure key using two sets of
bases that were mutually unbiased.  Later, Ekert suggested the use
of entangled states to generate a common key in a secure fashion
\cite{Ek1}. However, these quantum key distribution (QKD)
protocols involved only two parties and two-level systems. In
recent years researchers have drawn their attention to QKD
protocols that involve multi-level systems with two parties
\cite{Durt1,Bech1,Moh1,Moh2,cerf1,Dag1}, or multiple parties with
two-level systems \cite{Mark1,Scar1}. Motivating the pursuit of
multi-level quantum key distribution is that more information can
be carried by each particle thereby increasing the information
flux, and some multi-level protocols have been shown to have
greater security against eavesdropping attacks \cite{Durt1,Dag1}.
As for multi-party protocols there is the quantum secret sharing
protocol which employs $|GHZ\rangle _{N}$ states
\cite{Mark1,Scar1}, but there seems to be little else besides
this.

    On the experimental aspect, one of the obstacles for multi-level
schemes is the feasibility of such schemes. Atoms have multiple
energy levels that can be utilized, but preparing atoms in some
prescribed state and sending them off to separate parties is not
realistic.  The decoherence time of the state will determine how
far the particles can travel before they become useless for any
scheme that requires a particular state. However, recent
experimental demonstrations in the entanglement of orbital angular
momentum states of photons and the generation of arbitrary
entangled states with these orbital angular momentum quantum
numbers \cite{Mair1,Torres1,Sonja1} makes photons a promising
resource for multidimensional quantum protocols. Furthermore, much
work has been done in detecting these orbital angular momentum
states of light and its superpositions at the single photon level
\cite{Vaz1,Kot1,Leach1}.

    Here we investigate another possible multi-party protocol
involving a state which, unlike the $|GHZ\rangle _{N}$ state,
contains some entanglement even after one of the particles is
traced out. Although the remaining state is a mixed state,
perfectly correlated measurements can be made by making
measurements in a reduced space. This makes the state rather
interesting because it allows any two parties to create a key
without any help from the third.

\section{Quantum Secret Sharing Protocol}

    Suppose there is a task at hand where the involvement of more
than one party is needed for the sake of checks and balances. This
could be for launching missiles, opening bank safes, or other
sensitive matters where no one individual can be trusted to
execute. To this end, one sends only parts of the launch code,
bank vault combination, etc., to each party involved in the task.
The message can be deciphered only when all the parties involved
cooperate.  In recent years quantum mechanical version of these
secret sharing protocols have been discussed using GHZ states
\cite{Mark1,Scar1}.  Here we propose another secret sharing scheme
using a three-level system.

    We assume that the three parties (Alice, Bob, and Charlie) share the state

    \begin{equation}
|\Psi\rangle=\frac{1}{\sqrt{6}}\Bigg[\Big(|ab\rangle+|ba\rangle\Big)|c\rangle+
\Big(|ac\rangle+|ca\rangle\Big)|b \rangle
    +\Big(|cb\rangle+|bc\rangle\Big)|a\rangle\Bigg] \label{state}
\end{equation}

\noindent where $|a\rangle, |b\rangle$, and $|c\rangle$ are the
three quantum levels and
$\Big(|ab\rangle+|ba\rangle\Big)|c\rangle$ is short hand for
$\Big(|a\rangle _{\rm{Alice}}\otimes |b\rangle
_{\rm{Bob}}+|b\rangle _{\rm{Alice}}\otimes |a\rangle
_{\rm{Bob}}\Big)\otimes |c\rangle _{\rm{Charlie}}$. Note that this
state is the sum of all the permutations of the three levels, and
that the state collapses into a Bell state when one of the parties
makes a measurement in the representational basis hence the
measurement outcomes are perfectly correlated. Now we define
another set of measurement basis vectors

\begin{equation}
 |u1\rangle=\frac{1}{\sqrt{3}}\Big[|a\rangle+|b\rangle
+|c\rangle\Big], \label{mb1}
\end{equation}

\begin{equation}
 |u2\rangle=\frac{1}{\sqrt{3}}\Big[|a\rangle+e^{i\phi}|b\rangle
+e^{-i\phi}|c\rangle\Big], \label{mb2}
\end{equation}

\begin{equation}
 |u3\rangle=\frac{1}{\sqrt{3}}\Big[|a\rangle+e^{-i\phi}|b\rangle
+e^{i\phi}|c\rangle\Big], \label{mb3}
\end{equation}

\noindent where $\phi=\frac{i2\pi}{3}$.  This set of measurement
basis vectors is a mutually unbiased basis set for a three level
system. The original state is perfectly correlated in this
measurement basis as well since

\begin{equation}
\langle u1,u1|\Psi\rangle=|u1\rangle, \langle
u2,u1|\Psi\rangle=-|u2\rangle, \langle
u3,u1|\Psi\rangle=-|u3\rangle,
\end{equation}

\begin{equation}
\langle u1,u2|\Psi\rangle=-e^{-i\phi}|u3\rangle,  \langle
u2,u2|\Psi\rangle=-e^{-i\phi}|u1\rangle,  \langle
u3,u2|\Psi\rangle=e^{-i\phi}|u2\rangle,
\end{equation}

\begin{equation}
\langle u1,u3|\Psi\rangle=-e^{i\phi}|u2\rangle,  \langle
u2,u3|\Psi\rangle=e^{i\phi}|u3\rangle,  \langle
u3,u3|\Psi\rangle=-e^{-i\phi}|u1\rangle
\end{equation}

\noindent where $\langle u1,u1|\Psi\rangle=|u1\rangle$ is
shorthand for $_{\rm{Bob}}\langle u1|\otimes _{\rm{Alice}}\langle
u1|\Psi\rangle=|u1\rangle_{\rm{Charlie}}$.  First, Alice measures
her particle using one of the bases, then Bob makes his
measurement in one of the bases and then Charlie does the same. If
all the parties involved measure in the same basis, then they will
keep the outcome of their measurement. At the very end, Bob and
Charlie get together and compare notes to determine Alice's
measurement outcomes. Clearly, from the structure of the initial
state, neither Bob nor Charlie could tell what Alice's measurement
was without getting together and sharing measurement results.

\section{Quantum Key Distribution Protocol}

Alice, Bob, and Charlie still share the same initial state
described before, but what happens if Charlie loses his particle?
Can Alice and Bob still utilize the entanglement they have between
their particles to communicate?  There is indeed a simple way to
take advantage of the residual entanglement Alice and Bob share.
The reduced density matrix of the original state when Charlie's
system is traced out is

\begin{equation}
\hat{\rho}_{AB}=\frac{1}{3}\Big[|\Psi_{ab}\rangle\langle\Psi_{ab}|+|\Psi_{bc}\rangle\langle\Psi_{bc}|+|\Psi_{ca}\rangle\langle\Psi_{ca}|\Big]
\end{equation}

\noindent where
$|\Psi_{ij}\rangle=\frac{1}{\sqrt{2}}[|ij\rangle+|ji\rangle] $ and
$i,j\in(a,b,c)$.  Alice and Bob share this mixed state, but the
question remains whether they can get perfectly correlated
measurement outcomes from this state.  Indeed, this can be done if
Alice and Bob restrict their measurements to a two-dimensional
subspace of the three-level system.

Let us supposed Alice and Bob decide to make measurements in the
$\big(|a\rangle , |b\rangle\big)$ subspace, so they measure in
either the $\Big\{|a\rangle, |b\rangle\Big\}$ basis or
$\Big\{\frac{1}{\sqrt{2}}\Big(|a\rangle+|b\rangle\Big),\frac{1}{\sqrt{2}}\Big(|a\rangle
-|b\rangle\Big)\Big\}$ basis.  If the state they shared was
$|\Psi_{ab}\rangle$, then they would get perfectly correlated
measurement outcomes provided they measured in the same basis.  In
the case in which the state they shared was $|\Psi_{bc}\rangle$
either Alice or Bob will get a click in his or her detector if
they measure in the $\{|a\rangle, |b\rangle\}$ basis since
$|\Psi_{bc}\rangle$ has a component in $|b\rangle$. However, in
this case it is impossible for both Alice and Bob to get a click
in their detectors, since if one measures the state of the
particle to be in $|a\rangle$, then the other party's particle
will be in state $|c\rangle$, which is not within the
two-dimensional subspace in which they are making the measurement.
A similar argument holds for the
$\Big\{\frac{1}{\sqrt{2}}\Big(|a\rangle+|b\rangle\Big),\frac{1}{\sqrt{2}}\Big(|a\rangle
-|b\rangle\Big)\Big\}$ basis, it is impossible for both Alice and
Bob to get a click in their detectors.  Hence, for QKD purposes
Alice and Bob will disregard the measurements in which: 1) they
did not measure in the same basis, and 2) when they did not both
register a click in their detectors. The remaining measurements
they made will be perfectly correlated.

In fact, Alice and Bob don't even need to previously agree upon
the subspace in which they make the measurement.  They can
randomly choose the subspace and add to the two previous criteria
that they also disregard the measurements made in different
subspaces.

\section{Realization Using Orbital Angular Momentum of Light}

Although the protocol is independent of any particular
realization, here we present an implementation of the protocol
using orbital angular momentum states of light.  We present both a
method to generate the initial entangled state, and the means to
detect both the orbital angular momentum states and its
superposition.

It has been experimentally verified that the orbital angular
momentum of a photon is conserved through spontaneous parametric
down conversion, and the daughter photons are entangled in their
orbital angular momentum \cite{Mair1}. Since there is no upper
bound to the orbital angular momentum a photon can carry, it is
ideal for multidimensional quantum protocols.

First, we will have to generate the state the three parties are
going to share.  Here we will use three entangled sources, a three
beam coupler, three detectors, and a computer hologram to
differentiate between the different orbital angular momentum
states of the photon.  The method used is in the same spirit as
the method used to generate GHZ states from two entangled sources
\cite{Anton1}.

The entangled source of light we are going to use is generated
through spontaneous parametric down conversion.  Using a suitable
computer generated hologram to modify the pump beam, we can
produce the following orbital angular momentum entangled state
\cite{Torres1},

\begin{equation}
|\Psi_{\rm{source}}\rangle=\frac{1}{\sqrt{3}}\Big[|0,0\rangle
+|1,1\rangle +|2,2\rangle\Big].
\end{equation}

\begin{figure}
\includegraphics[width=3.3in]{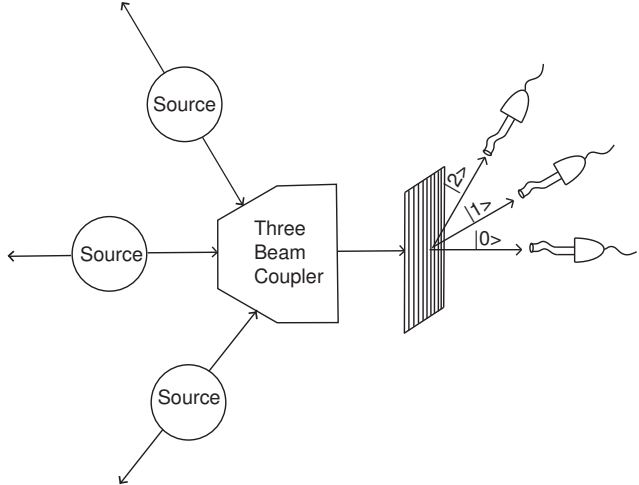}
\caption{\label{source} Generation of tripartite three-level
entangled state. The three photons that do not get detected are in
the state $|\Psi_{source}\rangle$ provided all three detectors
detect a photon}
\end{figure}

\noindent We then take three of these sources and send one of each
source's output into a three-beam coupler.  At the output of the
coupler we put another computer generated hologram with one
dislocation and we place a single mode fiber that goes into a
detector at each of the three diffraction orders as shown in Fig.
\ref{source}. The hologram imparts a $\Delta l=0$ for the zeroth
diffraction order, $\Delta l=1$ for the first diffraction order,
$\Delta l=2$ for the second diffraction order, and so on to the
input beam.  The single mode fibers only couple in the lowest
order orbital angular momentum states hence the detector placed in
the second diffraction order will only click if the diffracted
photon was originally in the $l=2$ state \cite{Mair1}.  If all
three detectors register a photon then it means that the photons
that weren't detected have orbital angular momentum of $l=0, l=1,$
and $l=2$, but we do not know which photon carries which state.
Hence we are left with the state

\begin{equation}
|\Psi_{\rm{tripartite}}\rangle=\frac{1}{\sqrt{6}}\Big[|0,2,1\rangle+|0,1,2\rangle+|1,0,2\rangle+|1,2,0\rangle+|2,0,1\rangle+|2,1,0\rangle\Big]\nonumber
\end{equation}
\begin{equation}
=\frac{1}{\sqrt{6}}\Bigg[\Big(|2,1\rangle+|1,2\rangle\Big)|0\rangle+\Big(|2,0\rangle+|0,2\rangle\Big)|1\rangle+\Big(|0,1\rangle+|1,0\rangle\Big)|2\rangle\Bigg].
\end{equation}

\noindent This is the original state with which we started, Eq.
(\ref{state}), by replacing $|a\rangle, |b\rangle,$ and
$|c\rangle$ with $|0\rangle, |1\rangle,$ and $|2\rangle$.

Now that we have the state which the three parties share, the
problem we are left with is to detect the orbital angular momentum
states and its superposition.  This could also be done using
holograms \cite{Kot1,Vaz1}, but it is rather inefficient and it is
not particularly suitable when considering single photon states.
The method of choice here is a simple interferometric scheme
employing a Mach-Zehnder interferometer with Dove prisms in its
path \cite{Leach1}.

\begin{figure}
\includegraphics[width=3.3in]{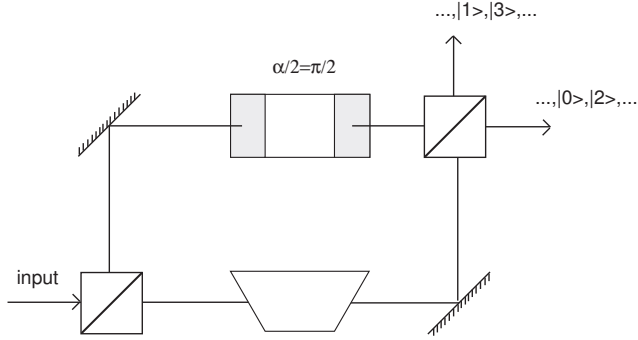}
\caption{\label{Dove} Sorting orbital angular momentum states of
light.  The Dove prisms in the two arms are rotated with respect
to one another by an angle of $\alpha/2=\pi/2$. With appropriate
path differences the even and odd orbital angular momentum states
emerge from different ports of the beam splitter.}
\end{figure}

In the first stage the Dove prisms in the two arms are rotated
with respect to one another by an angle of $\alpha/2=\pi/2$ (See
Fig. \ref{Dove}). This creates a relative phase shift between the
beams in the two arms of $\theta=l\pi$, where $l$ is the orbital
angular momentum quantum number.  The phase shift is produced
because the Dove prism flips the transverse structure of the
field.  Since the Laguerre Gaussian modes have a $e^{il\phi}$
phase structure, the Dove prism serves as a device that imparts a
$l$-dependent phase shift. Now, by adjusting the path difference
appropriately one can make it so that the odd and even orbital
angular momentum states come out of the two different output ports
of the interferometer.  The orbital angular momentum states of the
incoming beam can be sorted out by cascading these devices with
different angles between the Dove prisms \cite{Leach1}.  The
photon's state can then be collapsed into a particular $l$-state
by placing detectors at each of the output ports.

In detecting superposition states Eqs. (\ref{mb1}-\ref{mb3}), the
problem comes down to determining the relative phase difference
between the orbital angular momentum states.  Since orthogonal
states do not interfere with one another, we have to put holograms
at each output port of the sorting device to convert them all into
the same $l$-state. After this is done the photons are sent
through a three-port interferometer where the paths are
appropriately adjusted so that the three output ports are the
superposition states of interest \cite{Zuko1}.

For the case when only two of the three parties want to generate a
secure key the two parties use only two of the three output ports.
This too is easily done with the existing setup. After the sorting
device the two parties can measure in the orbital angular momentum
basis, or its superposition in the two-dimensional space.  Later,
they will divulge both their measurement basis and the subspace
they measured in to determine which measurements to keep.

\section{Conclusion}

Here we have shown a tripartite three-level system that can be
used for both secret sharing protocols involving all three
parties, or quantum key distribution protocol between any two
parties.  The two parties generate a secret key by taking
advantage of the residual entanglement of the reduced density
matrix.  This is done by making their measurements in a reduced
space.  A physical realization of this scheme has also been shown
through the use of entangled orbital angular momentum states of
photons.

\begin{acknowledgments}

We would like to thank John Howell, Govind Agrawal, Thomas Brown,
and Miguel Alonso for helpful discussions.  This work was
supported in part by the ARO-adminstered MURI Grant DAAD
19-99-1-0252.

\end{acknowledgments}

\end{document}